\documentstyle[11pt,newpasp,twoside,epsf]{article}
\makeindex
\index{a:Duc, P.--A.}
\index{a:Bournaud, F.}
\index{a:Masset, F.S.}

\index{k:Galaxies-general!dark matter}
\index{k:Galaxies-general!dwarf} 
\index{k:Interactions!numerical simulations} 
\index{k:Interstellar medium!metal enrichment} 
\index{k:Tidal Dwarf Galaxies!census} 
\index{k:Tidal Dwarf Galaxies!kinematics}
\index{k:Tidal Dwarf Galaxies!dark matter}
\index{k:Tidal Dwarf Galaxies!metallicity}

\def\edcomment#1{\iffalse\marginpar{\raggedright\sl#1\/}\else\relax\fi}
\reversemarginpar

\begin{document}
\title{Identifying old Tidal Dwarf Galaxies in Simulations and in the Nearby Universe}
\author{Pierre--Alain Duc$^1$, Fr\'ed\'eric Bournaud$^{1,2}$, Fr\'ed\'eric Masset$^1$} \affil{$^1$
CNRS FRE 2591 and  Service d'astrophysique, CEA--Saclay, France \\ $^2$ LERMA, Observatoire de Paris, France}

\begin{abstract}
Most Tidal Dwarf Galaxies (TDGs) so-far discussed in the
literature may be considered as young ones or even newborns, as
they are still physically  linked to their parent galaxies by an
umbilical cord: the tidal tail at the tip of which they are
usually observed. Old Tidal Dwarf Galaxies, completely detached
from their progenitors, are still to be found. Using N--body
numerical
 simulations, we have shown that tidal objects as massive  as $10^{9}$ solar masses may be
formed in  interacting systems and survive for more than one Gyr. Old TDGs should hence
exist in the Universe. They may be identified looking at a peculiarity  of their  ``genetic identity card":
a relatively high abundance in heavy elements, inherited from their parent galaxies.
Finally, using this technique, we revisit the dwarf galaxies in the local Universe trying to find
arguments pro and con a tidal origin.
\noindent
\end{abstract}

\section{Introduction}
The presence of compact star-forming regions in the tidal tails of
colliding galaxies is commonly observed. The formation, in that
environment,   of super-star clusters with  masses up to those of
globular clusters has also often been reported (e.g., Knierman et
al., 2003). The existence of even more massive tidal objects, with
global properties characteristics of dwarf galaxies, has been
claimed for more than a decade (see the  review by Duc \& Mirabel,
1999). But whether such ``Tidal Dwarf Galaxies" are genuine
galaxies -- i.e.  they are  gravitationally bound entities --
 is still strongly debated (see in this volume the contribution by Hibbard 
\& Barnes  and  that of Amram et al. and Braine et al. 
for an alternative view). But  perhaps the more important issue of whether  TDGs survive, escape from
their parent galaxies, and significantly  contribute to the overall population of dwarf galaxies, is  a matter
of speculation. Indeed, a more  plausible fate
is their tidal destruction or their falling back onto their progenitors. There are two ways to tackle
the problem: using numerical simulations of  interacting galaxies and following in them the
buildup and evolution of tidal objects, or trying to identify in catalogs of  dwarf  galaxies  those which might have a
tidal origin. We have explored both approaches.

\section{Old Tidal Dwarf Galaxies in numerical simulations}
Since the seminal work by Toomre \& Toomre (1972), numerous
numerical simulations have been made to study the evolution of
colliding galaxies. If the reliability of the computer
calculations has often been checked from their ability to produce
realistic  tidal tails, most of these efforts were focussed in
understanding what happens in the most central regions. The
simulations by Barnes \& Hernquist (1992) were the first ones to
show  the formation, out of tidal material,  of bound clumps with
masses of up to $10^{8} M_{\odot}$, distributed all along the
tails. As stressed later on, among others by Hibbard \& Mihos
(1995), the tidal material will quickly fall back unless it was
originally sent to large distances where it may survive for more
than a Hubble time. In real interacting systems, however, this is
precisely where the most massive TDG candidates, with apparent
masses of a few  $10^{9} M_{\odot}$, are found (unless these are
just the result of projection effects; see arguments against that
idea in the contribution by Amram et al. in this volume). Very
recently, we were able to reproduce for the first time in N-body
simulations the formation of such massive objects near the tip of
long tidal tails (Bournaud, Duc \& Masset, 2003). This could be
achieved when  we  adopted very extended dark matter halos for
both parent galaxies. Such large halos are actually consistent
with theoretical cosmological models; they are however usually
disregarded and truncated in numerical simulations because they
consume a large number of  particles (and CPU time) but have only
a minor impact on the merger remnant where most of the attention
is. Figure~1 presents our simulations; it shows how the gas of one
galaxy with an extended dark matter halo reacts to the
perturbation of another one of the same mass. Note in particular
how the gas piles up at the tip of one of the tidal tails.
Self-gravity makes it become a compact object with a mass of
$10^{9} M_{\odot}$.  Two Gyr after the first encounter and the
dissolution of the tails, it is still visible on a quasi-circular
orbit at large radii and appears as a classical companion galaxy.
In simulations where long tidal tails are able to form (and hence
for which the colliding galaxies have the adequate orbital
parameters and relative velocities), the production of massive,
long-lived,
 tidal objects is not exceptional. However, we still need to explore more systematically the parameter
space to obtain quantitative predictions on the production rate of TDGs.
In any case, our preliminary study indicates that massive tidal objects may be created during
tidal collisions, and have a life time long enough to become  old Tidal Dwarf Galaxies, as
previously defined. Therefore searching for such objects in the real Universe makes sense.

\begin{figure}[ht!]
\plotone{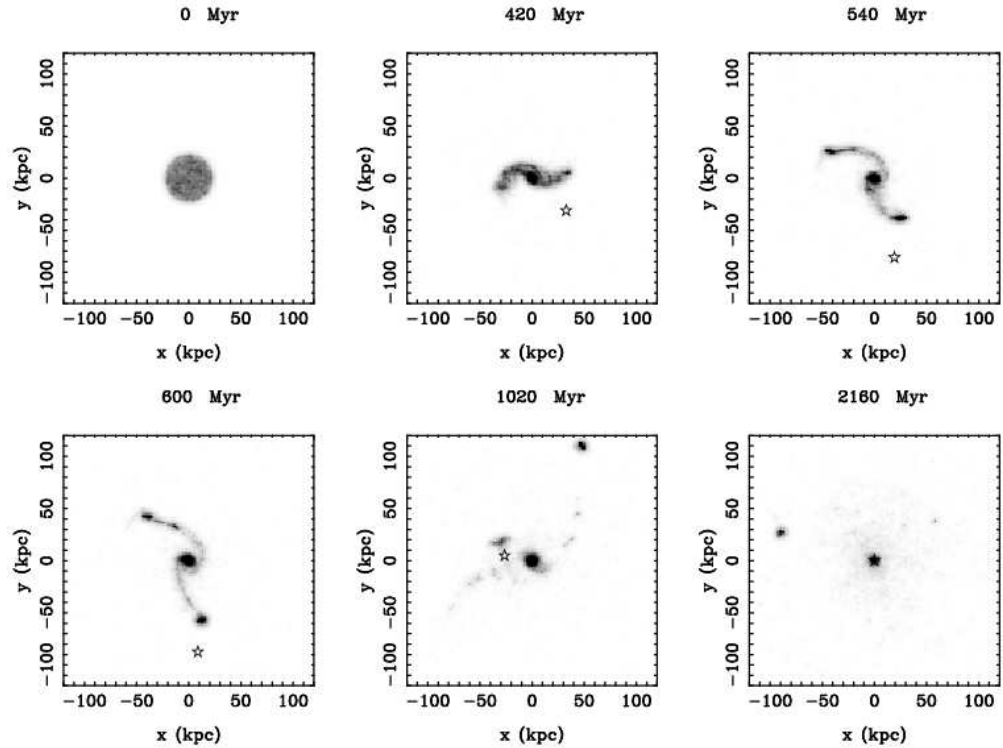} \caption{Response of the gaseous disk of a galaxy to a tidal interaction. The gas
column-density is plotted in grayscale. The center of the second galaxy is represented by the star symbol.
The truncation radius of the dark haloes is extended up to  10 times the radius of the stellar disks
(see details in Bournaud, Duc \& Masset, 2003)
}
\end{figure}

\section{Identifying old Tidal Dwarf Galaxies}
How to identify old TDGs when they have no obvious connection -- an optical stellar  bridge or an HI
gaseous bridge -- to any nearby merging galaxy ?
Hunter et al. (2000)  proposed to look at their dark-matter content, expected to be small if TDGs are made
of disk material and if most of the dark matter is distributed in a halo. Such objects should also have a special
location on the Tully-Fisher diagram. However, by far the easiest way is an unusually high  metallicity.
Standard isolated galaxies follow a fair correlation between their luminosity (mass) and their
metallicity. The less massive ones tend to be less metal-rich because they are less able to
retain their heavy elements that the most massive ones. This correlation is shown in Figure~2.
Galaxies departing form that relation and having  an oxygen abundance which is too high
for their luminosity can be considered as TDG candidates. Indeed tidal dwarfs are made of recycled,
pre-enriched, material. In fact, the oxygen abundances of young tidal objects (observed near
confirmed mergers, and shown in Fig.~2) have an oxygen abundance of about 1/2 - 1/4 solar, which
is independent  of their luminosity, and typical of that of the external regions of spiral disks from where
their building material originally comes.

\begin{figure}[ht!]
\plotone{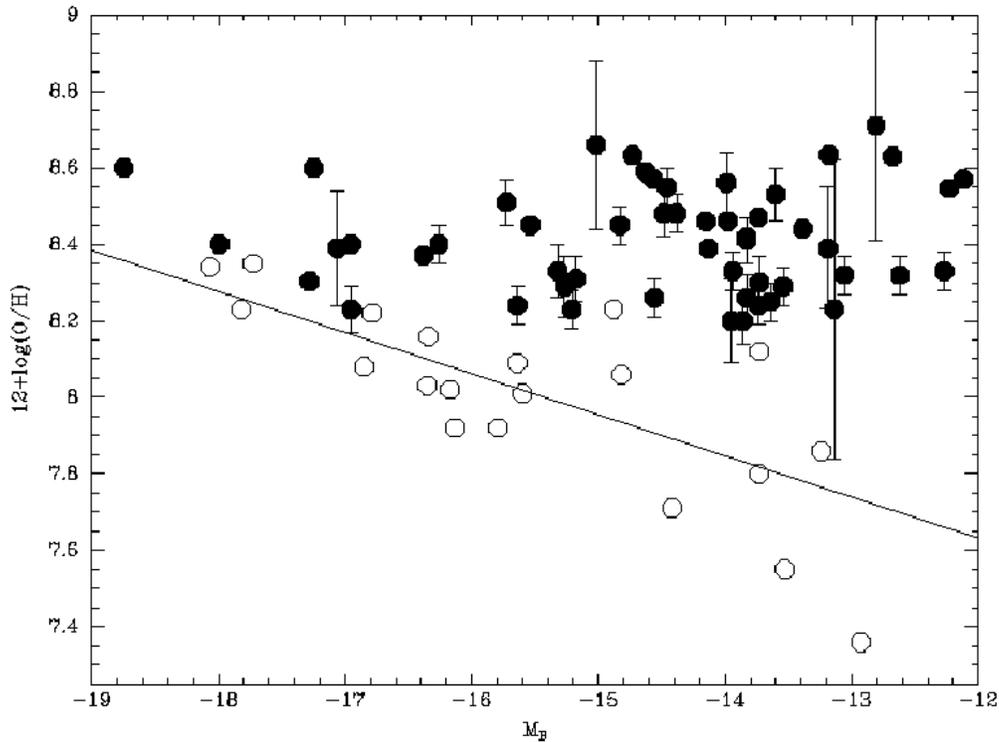} \caption{Luminosity-metallicity relation for a sample of isolated nearby
dwarf galaxies (open circles; Richer \& McCall, 1995) and in the HII regions of a sample of
young tidal objects (black circles; Weilbacher et al, 2002). The oxygen abundances versus
absolute blue magnitudes are plotted.}
\end{figure}

One should note that other phenomena can account for    a high metallicity in a dwarf galaxy .
An external confinement effect by a dense intergalactic medium may prevent it
from expelling its heavy elements. Some dwarfs could also be the remnants of more massive
 and metal--rich galaxies that have been disrupted by collisions (see the contribution of K. Bekki
in this volume). Those effects should principally apply in dense environments such as
in clusters of galaxies.

As a side effect, molecular gas as traced by the  CO emission
should be more easily detectable in pre-enriched dwarfs whereas it
is not observed in the metal poor classical dwarfs. Strong CO
emission has been detected in most massive TDGs (Braine et al.
2001, and his contribution in this volume).

\section{Old Tidal Dwarf Galaxies in the nearby Universe}
Is there any observational evidence that TDGs may survive for at least 500 Myr and
may hence already  be considered as old ? First of all, a few confirmed TDGs  are observed
in the vicinity of advanced mergers. In these  systems, the nuclei of the colliding
galaxies have already merged. The associated time scale given by numerical simulations
is typically 0.5-1 Gyr.  Examples of such objects are shown in Figure~3. In those, the
tidal tails are already quite faint whereas the most massive and distant condensations
are still clearly visible. Optical spectroscopy indicates that all of them have oxygen abundances
consistent with those expected for a tidal origin.

\begin{figure}[ht!]
\plotone{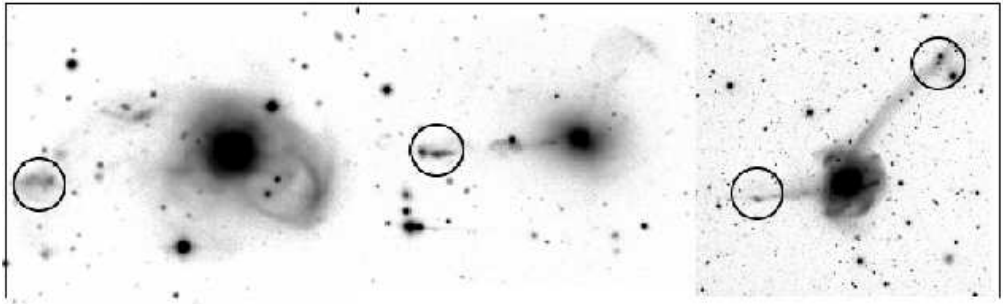} \caption{Tidal Dwarf Galaxies (encircled) in the vicinity of three advanced mergers.
The dynamical age of  these systems are  at least 500 Myr.}
\end{figure}

\index{o:Local Group}
\index{o:Sagittarius dwarf} 
\index{o:M 81 group}
\index{o:Garland}
\index{o:Fornax cluster}
\index{o:Coma cluster}
\index{o:Virgo cluster}
\index{o:Hydra cluster}

We explore in the following whether even older, completely detached, TDGs are present
 in catalogs of classical dwarf galaxies, starting with the Local Group.
Our closest  companion, the Sagittarius dwarf galaxy, is
currently being eaten by the Milky Way. Measuring with high-resolution UVES spectra element
abundances of a sample of individual stars, Bonifacio et al (2004) estimate that a substantial metal rich
( [Fe/H]=-0.25 ) population exists in Sgr. Such a high degree
of chemical enrichment either suggests that Sgr was a much larger galaxy in the past, or that
it was detached from the LMC or the MW during their interaction.
The distribution of the other dwarf spheroidals in the Local Group along a great circle
suggests that they also may have been involved in this collision (Lynden-Bell, 1982). However, their
metallicities, usually considered to be low  and their
high dark matter content (although challenged by Kroupa, 1997) make them unlikely TDG candidates.

One of the closest groups of galaxies, the M81 group, is well
known for the  spectacular tidal interaction in which  three of
its members are involved: M81, M82 and NGC 3077. Numerous TDG
candidates or  intergalactic HII regions  were identified in that
environment (Makarova et al., 2002;  Durrell et al. in this
volume);  among them
 The Garland object (Walter et al., in this volume),
and the protogalactic molecular cloud near Holm IX  are best studied.
Whereas,  on optical images, the latter intergalactic objects appear "detached" ,  HI maps show that
they actually lie  towards the prominent gaseous tidal tail linking the three main galaxies and in which they
were probably born. They cannot hence be considered as old TDGs.

The closest cluster of galaxies, Virgo, has its own TDG candidate: a HI, CO and Oxygen rich
low-surface brightness galaxy, not too far from the lenticular NGC 4694 (Braine et al.,  in
this volume). Studying the early-type dwarf galaxies in the nearby clusters of Coma and
Fornax, resp. Rakos et al. (2000) and Poggianti et al. (2001; see also her contribution in this volume),
found that a significant fraction of them  deviate from the metallicity-luminosity relation;
the youngest ones being the more metal-rich ones.
We reached  the same conclusion in our own complete survey of HI-rich, star--forming dwarf
galaxies in the Hydra (Duc et al., 2001) and Hercules (Iglesias-Paramo et al., 2003) clusters:
about 30\% of them have unusually high oxygen abundances in their HII regions.

It is still premature to conclude from these surveys whether old
TDGs contribute significantly to the population of dwarf galaxies.
As already mentioned, other phenomena may account for the
deviating metallicities.  On the other hand, selection criteria
only based on  metallicities disregard TDGs that would have formed
out of the most external and metal--poor disk material, or those
formed in the early Universe when collisions were more frequent
and galaxies less metal--rich. Likely the frequency of TDGs will
depend on the environment. Compact groups may be a particularly
favorable one (Hunsberger et al., 1996). Our numerical simulations
indicate that long-lived (i.e., formed at least 1-2 Gyr ago) TDGs
should exist. Whether they will survive for a Hubble time is still
unknown. One may  however note that the falling back onto a giant
galaxy is also the fate of numerous classical dwarfs.

\end{document}